\documentclass[twocolumn,preprintnumbers,amsmath,amssymb,prl]{revtex4}

\usepackage{graphicx}
\usepackage{dcolumn}

\usepackage{mathptmx, courier, pifont}
\usepackage[scaled=0.92]{helvet}
\usepackage[T1]{fontenc}
\usepackage{textcomp}

\usepackage{bm}

\newcommand{\singlefig}[6]{%
\begin{figure}\vspace{#3}%
\includegraphics*[scale=#5]{#2}%
\caption{\label{fig:#1} #6}%
\vspace{#4}%
\end{figure}}

\newcommand{\doublefig}[6]{%
\begin{figure*} \vspace{#3}%
\includegraphics*[scale=#5]{#2}%
\caption{\label{fig:#1} #6}
\vspace{#4}
\end{figure*}}

\newcommand{\tsub}[1]{_{\mbox{\scriptsize#1}}}

\newcommand{\sufourk}{SU(4)$_k$}

\begin{document}

\title
{ The Origin of Fermi Arcs in Cuprate Pseudogap States and Strong
Constraints on Viable Theories of High-Temperature Superconductivity
}

\author
{ Mike Guidry$^{(1)}$, Yang Sun$^{(2)}$, and Cheng-Li Wu$^{(3)}$ }
\affiliation{ $^{(1)}$Department of Physics and Astronomy,
University of Tennessee, Knoxville, Tennessee 37996--1200, USA
\\
$^{(2)}$Department of Physics, Shanghai Jiao Tong University,
Shanghai 200240, People's Republic of China
\\
$^{(3)}$Physics Department, Chung Yuan Christian University,
Chung-Li, Taiwan 320, ROC
}

\begin{abstract}

A full Fermi surface  exists in underdoped high-temperature
superconductors if the temperature $T$ lies above the pseudogap
temperature $T^*$. Below $T^*$ only arcs of Fermi surface survive,
scaling with $T/T^*$ as $T \rightarrow  0$, with $T^*$ displaying
strong doping dependence.  There is no accepted explanation for this
behavior. We show that generalizing the BCS theory of normal
superconductivity to include  $d$-wave pairs and antiferromagnetism
leads to the origin and doping dependence of the $ T^*$ scale, and a
quantitative description of Fermi arcs. These results place strong
constraints on viable theories of high-temperature
superconductivity.

\end{abstract}

\date{\today}

\maketitle

High-temperature superconductors were discovered more than 20 years
ago \cite{highTc_discovery}, but there is no uniformly-accepted
explanation for their properties \cite{bonn06}. They exhibit
pseudogaps (PG) at lower hole doping: partial energy gaps occurring
below a temperature $T^*$ but above the superconducting critical
temperature $T\tsub c$ \cite{timu99}, with $T^*$ and $T\tsub c$
exhibiting strong and opposite doping dependence for low
hole-doping.  It is widely believed that understanding the PG states
and the $T^*$ scale is central to understanding the high-$T\tsub c$
mechanism \cite{bonn06,norm05}.

For normal superconductors---described well by
Bardeen--Cooper--Schrieffer (BCS) theory \cite{BCS57}---the  Fermi
surface (highest occupied fermion energy level) is key to
understanding superconductivity.  In optimally-doped and overdoped
high-$T\tsub c$ compounds a ``normal'' Fermi surface exists, and
BCS theory utilizing $d$-wave singlet hole pairs seems applicable
\cite{bonn06}.  In underdoped cuprates this picture fails:  PG
states (lying between $T^*$ and $T\tsub c$) have anomalous Fermi
surfaces, with spectral strength pushed away from the expected
Fermi energy by partial energy gaps.  This is termed ``gapping
out'',  or (more loosely) ``loss'', of the Fermi surface.

Angle-resolved photoemission spectroscopy (ARPES)
\cite{dama03,norm03} probes electronic properties
\cite{norm03,dag94} of states in high-temperature superconductors.
ARPES data suggest a decreased state density near the Fermi energy
for temperature $T<T^*$ that is anisotropic in momentum, with strong
$ T$-dependence.  A full Fermi surface is observed for $T >  T^*$;
as the temperature decreases below $ T^*$, only arcs centered on the
$d$-wave nodal lines (diagonals in momentum space) survive
\cite{norm98,zhou04,kani06}, scaling as $T/ T^*$ to zero length for
$T \rightarrow  0$ \cite{kani06}. Thus gapping  is  anisotropic in
momentum space, with an ungapped Fermi surface surviving only in the
form of temperature-dependent {\em Fermi arcs}.

Earlier theoretical work on Fermi arcs has employed a variety of
approaches, including precursor pairs \cite{enge98,preo99}, the
perturbative renormalization group  \cite{furu98}, high-temperature
expansions in the $t$-$J$ model \cite{puti98}, RVB models with
strong gauge coupling \cite{wen98}, and time-reversal violating
phases \cite{varm06}, but there is no agreed-upon explanation of
Fermi arcs. The issue received renewed focus because of results  by
Kanigel, et al \cite{kani06} that place the strongest constraints
yet on the nature of Fermi arcs. We report here a quantitative
description of the Fermi-arc data in Ref.\ \cite{kani06}, with a
theory that also consistently predicts the pseudogap temperature
$T^*$  at the heart of the Fermi-arc scaling behavior. Thus, we
offer a comprehensive solution to one of the biggest mysteries in
high-temperature superconductivity:  the origin of the pseudgap and
of Fermi arcs in underdoped cuprates. More importantly, our results
imply that only high-$T\tsub c$ models with a strongly-constrained
set of properties can be consistent with the detailed attributes of
Fermi arcs and the scale $T^*$.

In earlier work we proposed an SU(4) model for the ground state
properties of cuprate systems. Generalized SU(4) coherent states are
the simplest Hartree--Fock--Bogoliubov variational solutions that
incorporate antiferromagnetism (AF) competing with $d$-wave
superconductivity on a lattice with no double occupancy
\cite{guid99,guid04,sun05,sun06}. They generalize BCS  to
incorporate AF self-consistently. If AF correlations vanish, the
SU(4) gap equations reduce to the BCS gap equations; if instead
singlet pairing vanishes, the SU(4) gap equations describe an AF
spin system with N\'eel order; for finite AF and SC correlations,
the SU(4) gap equations describe the evolution with doping ($P$) and
temperature ($T$) of $d$-wave Cooper pairs interacting with strong
AF correlations.

SU(4) states forbid double occupancy by symmetry, not by projection
\cite{guid04}.  The exact zero-$T$ ground state at half filling is
an AF Mott insulator that evolves rapidly with hole doping into a
superconductor with strong AF correlations in the underdoped region,
and finally into a singlet, $d$-wave superconductor beyond a
critical hole doping $P_q \simeq\ 0.16$--$0.19$ \cite{sun05}. At
finite $T$ for  $P>P_q$, pairing is reduced by thermal fluctuations
and the pairing gap vanishes at $T_c$. In the underdoped region
($P<P_q$) there are quantum fluctuations associated with AF
correlations in addition to thermal fluctuations at finite $T$.  The
interplay between AF, pairing, and thermal fluctuations produces PG
states above $T\tsub c$ in which the fermionic pairs interact
through AF correlations.  Thus the SU(4) model identifies the energy
gap $\Delta_q$ opened by the AF correlations as the pseudogap, while
the (singlet and triplet fermion) pairs in these states may be
viewed as $d$-wave preformed pairs.

The SU(4) model can reproduce both the pseudogap and pairing gap
with doping dependence quantitatively in agreement with data (see
Ref.\ \cite{sun06}).  The SC transition temperatures $T\tsub c$  and
PG transition temperatures $T^*$ are also well reproduced, as
illustrated in Fig.\ \ref{fig:pseudogapPhaseDiagram}. This is our
first significant finding: SU(4) coherent states give descriptions
of pairing gaps and pseudogaps,  and their transition temperatures
$T\tsub c$ and $T^*$,  that are in quantitative agreement with
cuprate data.

\singlefig {pseudogapPhaseDiagram} {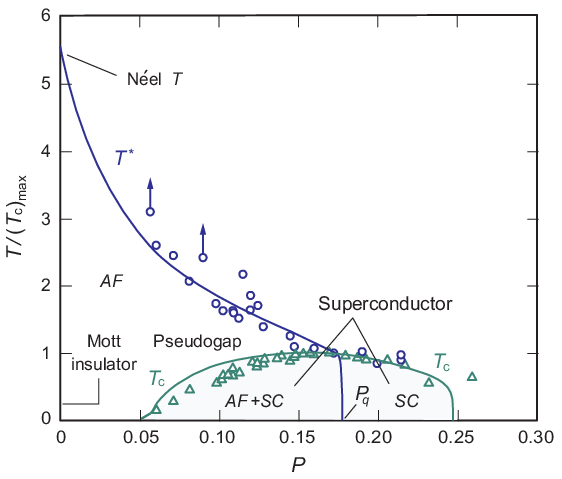}
{0pt} {0pt} {1.4} {(Color online) SU(4) cuprate   phase diagram
compared with data. Strengths of the AF and singlet pairing
correlations were determined by global fits to cuprate data
\cite{sun05,sun06}. The PG temperature is $T^*$ and the SC
transition temperature is $T\tsub c$.    The  AF correlations
vanish, leaving a pure singlet $d$-wave condensate, above the
critical doping  $P_q$. Dominant correlations in each region are
indicted by italic labels. Data from Ref.\ \cite{dai99} (arrows
indicate lower limits).}

\doublefig {gkPotentialComposite} {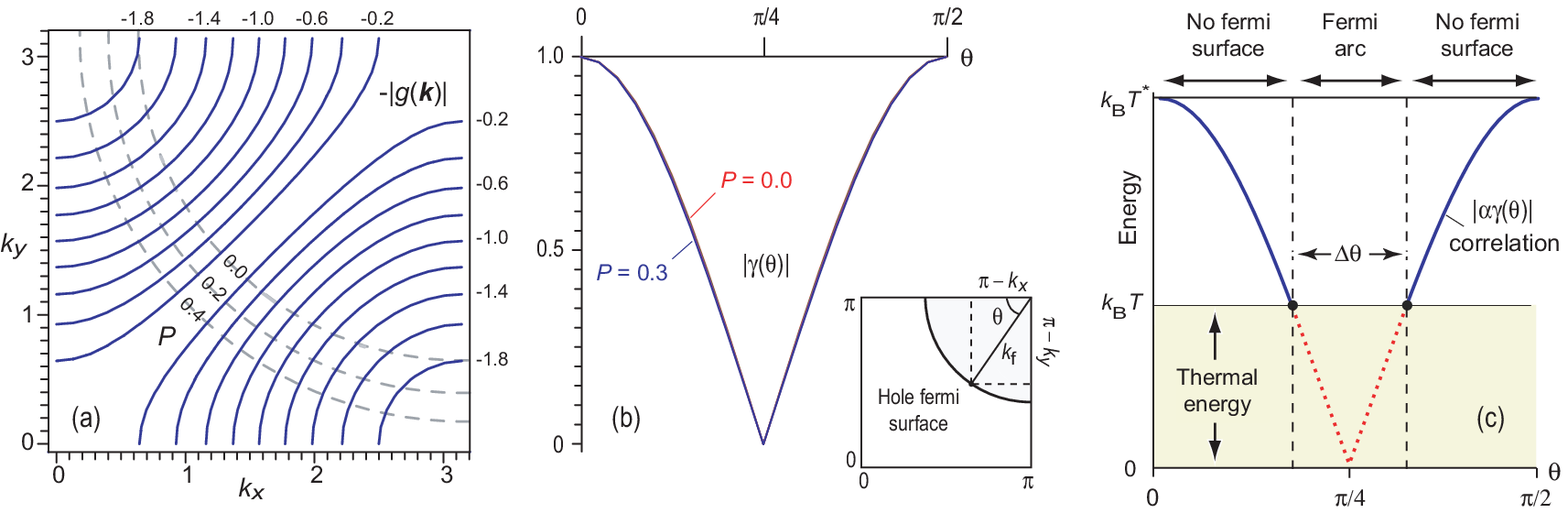}
{0pt} {0pt} {1.0} {(Color online) The $\bm k$-anisotropic factor,
pseudogap correlation energy, and graphical Fermi-arc solution.
(a)~The $\bm k$-anisotropic factor $|g(\bm k)|$ (blue lines) and
contours of equal hole doping $P$ (dashed gray lines),  in the
first Brillouin zone. (b) Correlation energy $\alpha
\gamma(\theta)$ for unit $\alpha=kT^*$ evaluated along the Fermi
surface [curves of constant $P$ in part (a)] as a function of the
angle $\theta$ (defined in inset).  Curves corresponding to doping
$P=$ 0--0.3 lie almost on top of each other, indicating that
$\gamma(\theta)$ is largely insensitive to doping. (c)~Graphical
solution of the Fermi-arc problem.  The curve defines the PG
correlation energy and horizontal lines correspond to constant
thermal energy scales $k\tsub B T$ associated with a given
temperature $T$.  Their intersections (black dots) represent
points in the momentum space where the pseudogap is just closed by
thermal fluctuations; these bracket arcs $\Delta\theta$ of
surviving Fermi surface. }

The 15 generators $\{{\cal G}_i \}$ of SU(4) are $M= \tfrac12
(n-\Omega)$ and
\begin{equation}
\begin{array}{ll}
   D^\dagger = \displaystyle
\sum_{\bm k} s_{gk} c^\dagger_{\bm k\uparrow} c^\dagger_{-\bm k\downarrow}
\quad
    \pi^\dagger_{ij} = \sum_{\bm k} s_{gk} c^\dagger_{\bm k+\bm q,i} c^\dagger_{-\bm k,j} \\[5pt]
 Q_{ij} = \displaystyle\sum_{\bm k} c^\dagger_{\bm k+\bm q,i} c_{\bm k,j}
\quad
  S_{ij} = \displaystyle\sum_{\bm k} c^\dagger_{\bm k,i} c_{\bm k,j} -\tfrac12\Omega \delta_{ij}
\end{array}
\label{pairOps1.1}
\end{equation}
and hermitian conjugates, where $\uparrow$, $\downarrow$, $i$, and
$j$ are spin indices, $\bm q \equiv (\pi,\pi,\pi)$, the maximum
number of doped holes or particles that can form coherent pairs
(assuming the half-filled normal state as vacuum) is $\Omega$, and
$s_{gk}$ is the algebraic sign of the $d$-wave pair formfactor
\begin{equation}
    g(\bm k) = g(k_x,k_y) = \cos k_x - \cos k_y .
\label{chi1.1}
\end{equation}
In these expressions $M$ is the charge with $n = \sum_{\bm k} n_{\bm
k}$ the electron number operator, and $\pi^\dagger_{ij}$, $Q_{ij}$,
and $S_{ij}$ are tensor components of the triplet pair, staggered
magnetization, and spin operators $\vec \pi$, $\vec Q$ and $\vec S$,
respectively.

These generators are summed over momentum $\bm k$.  They are
appropriate for describing data that are not momentum-selected (for
example, Fig.\ \ref{fig:pseudogapPhaseDiagram}).  However, ARPES
Fermi-arc data exhibit explicit dependence on $\bm k$.  The SU(4)
formalism may be extended to deal with this case by viewing the
individual $\bm k$ components of the operators defined in Eq.\
(\ref{pairOps1.1}) as the symmetry generators ${\cal G}(\bm k)$:
$$
\{{\cal G}(\bm k)\} \equiv \{D^\dagger(\bm k),D(\bm k),\vec{\pi}^\dagger(\bm k),\vec{\pi}(\bm k),\vec{Q}(\bm k),\vec{S}(\bm k),n_{\bm k}  \}
\label{kgenerators1.1}
$$
where, for example, the singlet pair generator $D^\dagger$ in Eq.\
(\ref{pairOps1.1}) is related to the $\bm k$-dependent generators
$D^\dagger(\bm k)$ by $ D^\dagger=\sum_{\bm k>0}D^\dagger(\bm k), $
and $\bm k>0$ means either $k_x>0$ or $k_y > 0$.

Instead of the global SU(4) symmetry generated by Eq.\
(\ref{pairOps1.1}), the symmetry now is a direct product of
$k$-dependent SU(4) groups, $\prod_{\bm k>0}\otimes$SU(4)$_{\bm k}$.
The corresponding Hamiltonian is
\begin{eqnarray}
    H &=& \sum_{\bm k>0}\epsilon_{\bm k}\,n_{\bm k}
    -\sum_{\bm k,\bm k'>0}\{\chi_{\bm k \bm k'}\vec{Q}(\bm k)\cdot\vec{Q}(\bm k')
\nonumber
\\
&& +G^{(0)}_{\bm k \bm k'}D^\dagger(\bm k)\,D(\bm k')
+G^{(1)}_{\bm k \bm k'}\vec{\pi}^\dagger(\bm k)\cdot\vec{\pi}(\bm k')
\} ,
\label{Hk1.1}
\end{eqnarray}
where $\epsilon_{\bm k}$ and $n_{\bm k}$ are single-particle
energies and occupation numbers, respectively, and the interaction
strengths are
$$
    \chi_{\bm k \bm k'}=\chi^0 |g(\bm k)g(\bm k')|
\qquad
G^{(i)}_{\bm k \bm k'}=G^{(i)}\delta_{\bm k} \delta_{\bm k'}\quad (i=0,1) ,
$$
where $\delta_{\bm k} \equiv  \delta(\theta) |g_0(\tilde k)|$. The
pair formfactor is expressed in terms of the magnitude of the hole
momentum $\tilde k$ and its azimuthal angle $\theta$: $g(\bm k)
\equiv g(k_x,k_y) = g(\tilde k,\theta)$, where $|g_0(\tilde k)|$ is
$|g(\tilde k,\theta)|$ at the antinodes ($\theta=0, \pi/2$), where
it is maximal, and $\delta(\theta)=1$ except in a very narrow region
near the nodal point, where it drops rapidly to zero at the node.

We shall term this $k$-dependent formalism the \sufourk\ model. All
results of the original $k$-independent SU(4) model of Refs.\
\cite{guid99,guid04,sun05,sun06} (including those of Fig.\
\ref{fig:pseudogapPhaseDiagram}) are recovered as a special case of
the \sufourk\ model for observables that are dominated by
contributions from near the Fermi surface ($\tilde k= k\tsub f$) and
averaged over $\bm k$ directions. However, general solutions of the
\sufourk\ model give new $k$-anisotropic properties.  Of direct
relevance to the present discussion is that the temperature for the
PG closure $T^*(\bm k)$ becomes anisotropic in $\bm k$;
specifically, we derive in the \sufourk\ model
\begin{equation}
    T^*(\bm k)\equiv T^*(k\tsub f, \theta) = T^*\left| g(k\tsub f,\theta)/g_0(k\tsub f) \right|
\label{TPG1.1}
\end{equation}
where $T^*$ is the gap closure temperature measured by ARPES along
the antinodal ($\theta = 0,\pi/2$) direction, which is the maximum
possible value of $T^*$ and should be somewhat larger than $T^*$
inferred from experiments that average over $\bm k$.
See  Ref.\ \cite{sun07}
 for details.
Equation (\ref{TPG1.1}) defines the full temperature and doping
dependence of Fermi arcs. The ultimate physical reason for this
result is that the singlet and triplet pairs interacting by AF
interactions in the SU(4) PG state each carry a $g(\bm k)$
formfactor, which introduces a $\bm k$ dependence in the effective
AF coupling and thus in $T^*$.

In Fig.\ \ref{fig:gkPotentialComposite}(a) we show the behavior of
$g(\bm k)$ as a function of ($k_x,k_y$). The components ($k_x,k_y$)
or $(k\tsub f,\theta)$ are constrained by the Fermi surface ($\tilde
k^2 = k^2_{\scriptstyle\rm f}$). Assuming an isotropic hole surface
[dashed gray lines in Fig.\ \ref{fig:gkPotentialComposite}(a)], we
have
\begin{equation}
    (\pi - k_x)^2 + (\pi-k_y)^2 = k_{\scriptstyle\rm f}^2 = 2\pi(1+P)
\label{arc1.3}
\end{equation}
(we consider a more realistic Fermi surface below). For a given
doping $P$ (thus $k\tsub f$) and temperature $T$, with $T^*(\bm
k)=T$ by virtue of  Eq.\ (\ref{TPG1.1}) under the  constraint
(\ref{arc1.3}), we obtain the angles  $\theta_1$ and $\theta_2$ at
which the PG closes (the angle $\theta$ is defined in the inset to
Fig.\ \ref{fig:gkPotentialComposite}(b)), and the length of the
surviving Fermi arc is $ k\tsub f |\theta_2-\theta_1| \equiv  k\tsub
f\Delta\theta$.

This result may be interpreted graphically.  In Fig.\
\ref{fig:gkPotentialComposite}(b) we show $\gamma(\theta) \equiv
|g(k\tsub f, \theta)/g_0(k\tsub f)|$ versus angle $\theta$ along
different Fermi surfaces [dashed gray lines in Fig.\
\ref{fig:gkPotentialComposite}(a)].  We see that $\gamma(\theta)$
is almost independent of doping $P$.  Thus, solving Eq.\
(\ref{TPG1.1}) with the Fermi surface constraint (\ref{arc1.3}) is
equivalent to solving
\begin{equation}
   k\tsub BT^*(\bm k)=\alpha\gamma(\theta) \qquad \alpha=k\tsub BT^* ,
\label{TPG1.2}
\end{equation}
with $k\tsub B$ the Boltzmann constant. The PG correlation energy
$\alpha \gamma(\theta)$ depends on the $\bm k$ direction $\theta$
[Fig.\ \ref{fig:gkPotentialComposite}(c)].  The pseudogap closes
when the thermal excitation energy $k\tsub B T$ is comparable to the
PG correlation energy.  The intersections of horizontal lines of
fixed $k\tsub B T$ with the correlation energy curve [black dots in
Fig.\ \ref{fig:gkPotentialComposite}(c)] define Fermi-arc solutions
$\theta_1$ and $\theta_2$.  Outside those points (solid blue curve),
the correlation energy is larger than the energy of thermal
fluctuations (shaded region), the PG opens, and the Fermi surface is
gapped.  Inside these points (dotted red curve), the thermal  energy
exceeds the correlation energy, the PG closes, and the Fermi surface
exists in an arc $\Delta\theta$ between the dots. When $T>T^*$, the
PG is closed in all directions and there is a full Fermi surface
since $k\tsub B T >\alpha$, and $\alpha$ is the maximum PG
correlation energy.

The arc solution exemplified graphically in Figs.\
\ref{fig:gkPotentialComposite}(b)--(c), or algebraically in Eq.\
(\ref{TPG1.1}), is our second significant finding: For given
temperature $T$, the anisotropy of  $\gamma(\theta)$ partitions
the $k$-space uniquely into regions having a Fermi surface
[$T>T^*(k\tsub f,\theta)$] and those that do not [$T<T^*(k\tsub
f,\theta)$]. As Fig.\ \ref{fig:gkPotentialComposite}(c) suggests,
arc lengths decrease with decreasing $T$ at fixed doping, with a
full Fermi surface at $T=T^*$, but only the nodal points at $T=0$.
Doping dependence enters primarily through the maximum PG
temperature $T^*$ and the Fermi momentum $k\tsub f$; the
temperature dependence enters through $T/T^*$.  For fixed $T$,
arcs shrink toward the nodal points with decreased doping because
$T^*$ in $T/T^*$ increases at smaller doping (Fig.\
\ref{fig:pseudogapPhaseDiagram}).  However, if Fermi-arc length is
measured by the fraction $\Delta\theta/(\pi/2)$ and the
temperature is scaled by $T^*$, the weak doping dependence of
$\gamma (\theta)$ ensures that $\Delta\theta/(\pi/2)$ versus
$T/T^*$ is almost independent of doping and compound.

Kanigel, et al \cite{kani06} report fractional arc lengths versus
$T/T^*$ for Bi2212 that we plot in Fig.\ \ref{fig:fermiArcLength}.
Our  theoretical solution  for the fractional arc lengths [obtained
from Fig.\ \ref{fig:gkPotentialComposite}(c) or Eq.\ (\ref{TPG1.1})]
is the solid line in Fig.\ \ref{fig:fermiArcLength}. Agreement
between data and theory is remarkable, given that the theoretical
curve has no adjustable parameters (it is determined completely by
the parameters fixed previously in Fig.\
\ref{fig:pseudogapPhaseDiagram}) and that we {\em predict} the scale
$T^*$ implicit in the data with the same theory (Fig.\
\ref{fig:pseudogapPhaseDiagram}). Note that the different behavior
at high and low temperatures in Fig.\ \ref{fig:fermiArcLength}
(rapid drop in arc length for decreasing $T\sim T^*$, shifting to
approximately linear decrease to zero arc-length at $T=0$), is
explained entirely by geometry  in Fig.\
\ref{fig:gkPotentialComposite}(c).

\singlefig
{fermiArcLength}                       
{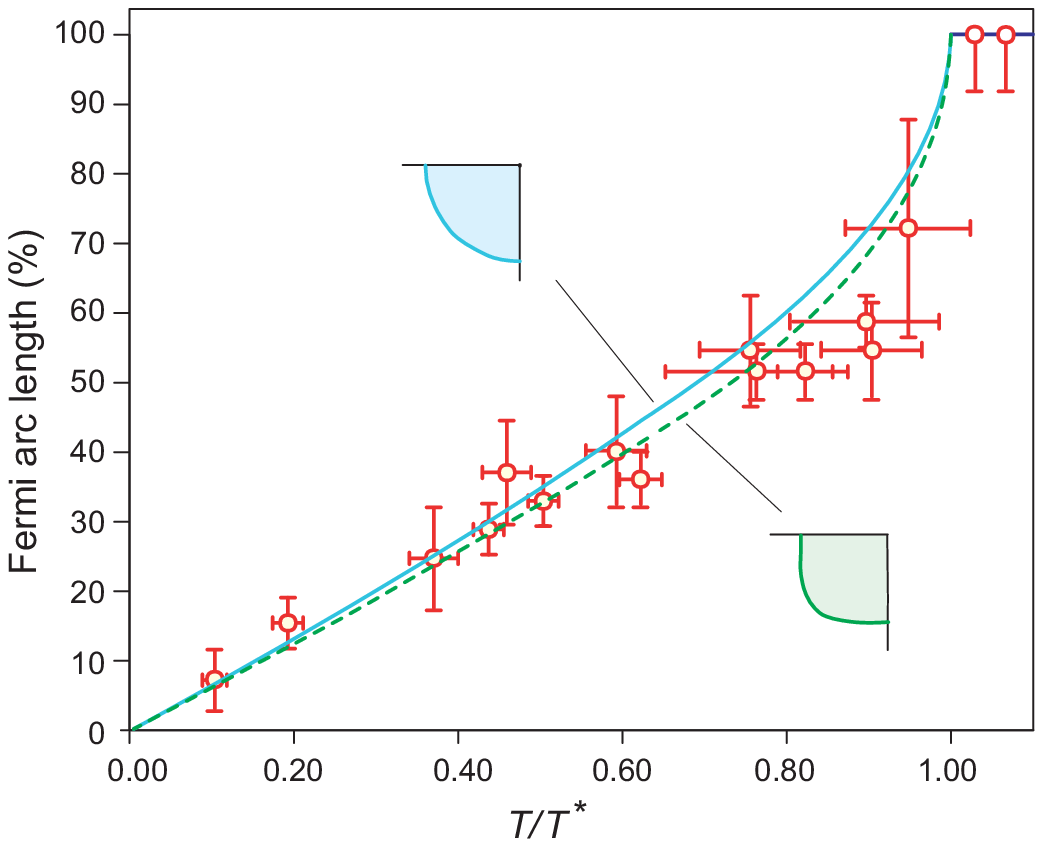}                   
{0pt}                          
{0pt}                          
{0.78}                          
{(Color online) Fermi arc length versus temperature. Experimental
arc length is displayed as a percentage of full Fermi surface
length vs.\ $T/T^*$ for underdoped Bi2212 \cite{kani06}; the solid
curve is our prediction for an isotropic Fermi surface (inset
upper left); the dashed curve assumes a Fermi surface with flat
antinodal segments (inset lower right). No parameters were
adjusted to these data in either calculation and the curves are
almost independent of doping. }

Realistic cuprate Fermi surfaces are flat near zone boundaries.
The dashed curve in Fig.\ \ref{fig:fermiArcLength} uses  the
flatter Fermi surface shown inset lower right. The similarity of
dashed and solid curves indicates that arc solutions depend only
weakly on differing curvature in nodal and antinodal regions.
Absolute arc lengths depend on doping and temperature.  However,
the scaled arc lengths of Fig.\ \ref{fig:fermiArcLength} {\em are
near-universal functions only of the ratio} $T/T^*$, largely
independent of compound, doping, and Fermi surface details, as
data suggest.

Figure \ref{fig:fermiArcLength} illustrates our third significant
finding: agreement between theory and data with no parameter
adjustment suggests that our SU(4) formalism outlines a complete
solution of the Fermi-arc mystery.  We now argue that this has
important implications for understanding pseudogap behavior.

The data in Fig.\ \ref{fig:fermiArcLength} have been interpreted
\cite{kani06} as representing a rapid drop from a full Fermi surface
at $T=T^*$, destroying the antinodal Fermi surface at essentially
constant temperature, followed by a linear decrease of Fermi-arc
length with decreasing $T/T^*$ on the near-nodal region of the Fermi
surface, extrapolating to zero arc length at the nodal points for
$T/T^* \rightarrow 0$.  This suggests that nodal and antinodal Fermi
surfaces may be removed differently as $T$ is lowered (see Ref.\
\cite{mcel06}).  For example, it has been argued \cite{kani06} that
abrupt removal of the antinodal surface at $T \simeq T^*$ may be
associated with a lattice vector connecting antinodal surfaces
(Ref.\ \cite{mcel06} speculates that gapping of the antinodal
surface may be associated with  charge ordering), while the
linearly-varying removal of the nodal regions extrapolating to a
nodal-point surface at $T=0$ may support a nodal liquid picture
\cite{bale98}.

The present results invite simpler hypotheses.  Our unified analysis
suggests that Fig.\ \ref{fig:fermiArcLength} is consistent with
removal of both nodal and antinodal surfaces by the {\em same
mechanism}.  Therefore, the qualitatively different variation of arc
length with $T/T^*$ in nodal and antinodal regions is not sufficient
to indicate that loss of Fermi surface proceeds by different
mechanisms in these regions. Nor do the data of Ref.\ \cite{kani06}
imply unique support for nodal liquids.  We have demonstrated
specifically that the SU(4) model accounts quantitatively for Fermi
arcs and $T^*$, without invoking nodal liquids.

We have shown that a  theory of Fermi arcs  must make two correct
predictions: (1)~the scale $T^*$ and its doping dependence, and
(2)~that $T^*(\bm k) \propto \gamma(\theta)$. {\em Any theory}
having a $T^*(\bm k)$  consistent with Eq.\ (\ref{TPG1.2}) can
describe the  scaled data of Fig.\ \ref{fig:fermiArcLength}, if
$T^*$ is taken from data. Hence, our fourth significant finding:
scaled ARPES data (Fig.\ \ref{fig:fermiArcLength}) test whether a
pseudogap has a nodal structure similar to that of the
superconducting state. But discriminating among different theories
meeting this condition requires more:  a quantitative,
self-consistent description of Fermi-surface loss and of the PG
temperature scale $T^*$ and its doping dependence (Fig.\
\ref{fig:pseudogapPhaseDiagram}). We conclude that only a
highly-restricted set of models can be consistent with the
aggregate properties of Fermi arcs. Finally, although our results
indicate that an SU(4) mean field can account for Fermi arcs, some
properties of the pseudogap state are expected to be influenced
significantly by SU(4) quantum fluctuations that we shall address
in forthcoming papers.

We thank Elbio Dagotto, Pengcheng Dai, and Takeshi Egami for
extensive discussion, and Elbio Dagotto for calling our attention to
an error in our original formulation.


\bibliographystyle{unsrt}

\end{document}